\magnification \magstep1
\raggedbottom
\openup 4\jot
\voffset6truemm
\headline={\ifnum\pageno=1\hfill\else
\hfill{\it Boundary operators in Euclidean 
quantum gravity} 
\hfill \fi}

\def\II{{\rm 1\!\hskip-1pt I}}

\def\sq{\Square}
\def\Square{\mathchoice{\square{8pt}}{\square{7pt}}{\square{6pt}}
    {\square{5pt}}}
\def\square#1{\mathop{\mkern0.5\thinmuskip
    \vbox{\hrule\hbox{\vrule
    \hskip#1 \vrule height#1 width 0pt \vrule}\hrule}\mkern0.5\thinmuskip}}

\def\cstok#1{\leavevmode\thinspace\hbox{\vrule\vtop{\vbox{\hrule\kern1pt
\hbox{\vphantom{\tt/}\thinspace{\tt#1}\thinspace}}
\kern1pt\hrule}\vrule}\thinspace}

\centerline {\bf BOUNDARY OPERATORS IN}
\centerline {\bf EUCLIDEAN QUANTUM GRAVITY} 
\vskip 1cm
\centerline {Ivan G Avramidi
\footnote{*}{Alexander von Humboldt Fellow. On leave of absence from Research 
Institute for Physics, Rostov State University,  Stachki 194, 
344104 Rostov-on-Don, Russia. 
E-mail: avramidi@math-inf.uni-greifswald.d400.de}, 
Giampiero Esposito
and Alexander Yu Kamenshchik} 
\vskip 0.3cm
\noindent
{\it Department of Mathematics, University of Greifswald,
Jahnstr. 15a, 17487 Greifswald, Germany}
\vskip 0.3cm
\noindent
{\it Istituto Nazionale di Fisica Nucleare,
Sezione di Napoli, Mostra d'Oltremare Padiglione 20,
80125 Napoli, Italy}
\vskip 0.3cm
\noindent
{\it Dipartimento di Scienze Fisiche, Mostra d'Oltremare
Padiglione 19, 80125 Napoli, Italy}
\vskip 0.3cm
\noindent
{\it Nuclear Safety Institute, Russian Academy of Sciences,
52 Bolshaya Tulskaya, Moscow 113191, Russia}
\vskip 0.3cm
\noindent
{\bf Abstract}. 
Gauge-invariant boundary conditions in Euclidean
quantum gravity can be obtained by setting to zero at the
boundary the spatial components of metric perturbations, and a
suitable class of gauge-averaging functionals. This paper shows
that, on choosing the de Donder functional, the resulting
boundary operator involves projection operators jointly with a
nilpotent operator. Moreover, the elliptic operator acting on
metric perturbations is symmetric. Other choices of
mixed boundary conditions, for which
the normal components of metric perturbations can be set to zero
at the boundary, are then analyzed in detail.
Last, the evaluation of the 1-loop divergence in the axial gauge
for gravity is obtained. Interestingly, such a divergence turns 
out to coincide with the one resulting from transverse-traceless
perturbations.
\vskip 1cm
\leftline {\bf 1. Introduction}
\vskip 0.3cm
\noindent
Over the last few years, a substantial progress has been
made in the understanding of the asymptotic heat kernel 
with pure and mixed boundary conditions in
quantum field theory. In particular, whenever the boundary
conditions for spinor fields, gauge fields and gravitation
are expressed in terms of complementary projection operators
[1], the geometric form of the 1-loop divergences is by
now well understood [2--4], and it agrees with the results 
obtained by analytic techniques [5--8]. What happens is that
the volume part of such 1-loop divergences involves the
curvature of the background, whilst the surface part
involves both the extrinsic and the intrinsic curvature
tensor of the boundary and the projection operators occurring
in the boundary conditions [2--4].

In Euclidean quantum gravity, however, a more general scheme
can be considered. As it has been shown in [8, 9], 
which rely on the work in [10],
one can set to zero at the boundary $\partial M$ 
the spatial components  
$h_{ij}$ of the metric perturbations $h_{ab}$,
jointly with any gauge-averaging functional 
$\Phi_{a}(h)$ which leads to a well defined spectrum of
eigenvalues (hereafter, lower-case indices $a,b$
should be regarded as abstract indices for four-dimensional tensor
fields). On requiring the invariance of such
boundary conditions under local transformations of metric
perturbations, i.e. $\delta h_{ab}=
\nabla_{(a} \; \varphi_{b)}$, one finds that a necessary and
sufficient condition for this is that the whole ghost 1-form
should vanish at the boundary [8, 9].
In particular, the background 
4-manifold $M$ can be taken to be flat Euclidean 
4-space bounded by a
3-sphere. The analysis of flat backgrounds is indeed relevant both
for Euclidean field theory [11] and for the analysis of 
massless supergravity models in the presence of boundaries [12].
In the de Donder gauge, 
the boundary conditions on the metric perturbations
which are invariant under gauge transformations
take then the form [8, 9]
$$
[h_{ij}]_{\partial M}=0
\eqno (1.1)
$$
$$
\left[{\partial h_{00}\over \partial \tau}
+{6\over \tau}h_{00}
-{\partial \over \partial \tau}\Bigr(g^{ij}h_{ij}\Bigr)
+{2\over \tau^{2}}h_{0i}^{\; \; \; \mid i}\right]_{\partial M}=0
\eqno (1.2)
$$
$$
\left[{\partial h_{0i}\over \partial \tau}
+{3\over \tau}h_{0i}-{1\over 2}h_{00|i}\right]_{\partial M}=0
\eqno (1.3)
$$
where $i,j=1,2,3$, and $g^{ij}$ are the spatial components of
the contravariant form of the background 4-metric. They are
used to raise indices of $h_{ij}$, whilst the covariant $g_{ij}$
lowers indices of $h^{ij}$. Moreover,
with a standard notation, 
$\tau=x^{0}$ is the radial coordinate,
${\hat x}^{i}$ are local coordinates on the 3-sphere boundary
with unit metric $c_{ij}(\hat x)$, so that locally
$$
g=d\tau \otimes d\tau + \tau^{2}c_{ij}(\hat x)
d{\hat x}^{i}\otimes d{\hat x}^{j} 
\eqno (1.4)
$$
and the stroke denotes covariant differentiation tangentially
with respect to the Levi-Civita connection of the boundary.
As we said before, the whole ghost 1-form should vanish
at the boundary.

Although the corresponding 1-loop divergence 
was already evaluated in [8]
by means of the regularization algorithm introduced in [13],
the geometric counterpart of such an analytic investigation 
remains unknown in the literature (cf [14]). 
Note that (1.2) and (1.3)
involve both normal and tangential derivatives of $h_{00}$ and
$h_{0i}$, and are not expressed in terms of (complementary)
projection operators. 

Thus, to complete the analysis of gauge-invariant boundary
conditions in Euclidean quantum gravity, it appears crucial to
perform a geometric analysis of the quantum boundary-value
problem corresponding to (1.1)--(1.3). For this purpose, section
2 describes the general framework for gauge-invariant boundary
conditions in Euclidean quantum gravity. Section 3 studies the
projection and nilpotent operators occurring in the de Donder
case. Section 4 obtains an equivalent form of the boundary
conditions of section 3, which makes it easier to compare them
with other sets of mixed boundary conditions studied in the
literature. Section 5 proves symmetry of the Laplace operator
when the boundary conditions (1.1)--(1.3) are imposed.
Section 6 is instead devoted to the analysis of
boundary operators when the normal components of metric 
perturbations are set to zero at the boundary. Section 7 evaluates
the 1-loop divergence for pure gravity in the axial gauge.
Concluding remarks are presented in section 8.
\vskip 0.3cm
\leftline {\bf 2. Gauge-invariant boundary conditions for
Euclidean quantum gravity}
\vskip 0.3cm
\noindent
For gauge fields and gravitation the boundary conditions should be 
gauge-invariant under local gauge transformations with some
suitable boundary conditions on the corresponding gauge functions
(ghost fields). This is why the boundary conditions
should be mixed, in that some components of the field
obey a set of boundary conditions (say, Dirichlet),
and the remaining part of the field obeys another
set of boundary conditions (Neumann or Robin).

We are here interested in the derivation of mixed boundary
conditions for Euclidean quantum gravity. The Euclidean 
formulation is essential to obtain well posed boundary-value
problems for elliptic operators. Its relevance for the
Lorentzian theory deserves further investigation [8],
since no general result holds which relates Lorentzian
and Riemannian curved four-manifolds through a Wick 
rotation, and the corresponding Green functions [11].
One can say, however, that our investigation of flat
backgrounds can be applied to put on solid ground the analysis
of ultraviolet divergences in quantum field theory on
manifolds with boundary.

The knowledge of
the classical variational problem, and the principle of
gauge invariance, are enough to lead to a highly non-trivial
quantum boundary-value problem. Indeed, it is by now well known
that, if one fixes the 3-metric at the boundary in 
general relativity, the corresponding variational problem
is well posed and leads to the Einstein equations, providing
the Einstein-Hilbert action is supplemented by a boundary
term whose integrand is proportional to the trace of the
second fundamental form [15]. In the corresponding
quantum boundary-value problem, which is relevant for the
1-loop approximation in quantum gravity, the perturbations
$h_{ij}$ of the induced 3-metric are set to zero at the
boundary. Moreover, the whole set of metric perturbations
$h_{ab}$ are subject to the infinitesimal {\it gauge}
transformations 
$$
{ }^{\varphi}h_{ab} \equiv h_{ab}+\delta h_{ab}
=h_{ab}+\nabla_{(a} \; \varphi_{b)}
\eqno (2.1)
$$
where $\nabla$ is the Levi-Civita connection of the background
4-geometry with metric $g$, and $\varphi_{a}dx^{a}$ 
is the ghost 1-form. In geometric language, the
infinitesimal difference between 
${ }^{\varphi}h_{ab}$ and
$h_{ab}$ is given by the Lie derivative along $\varphi$
of the 4-metric $g$. 

For problems with boundaries, equation (2.1) implies that
$$
{ }^{\varphi}h_{ij}=h_{ij}+\varphi_{(i \mid j)}
+K_{ij} \varphi_{0}
\eqno (2.2)
$$
where $K_{ij}$ is the extrinsic-curvature tensor of the boundary.
Of course, $\varphi_{0}$ and $\varphi_{i}$ are the normal
and tangential components of the ghost 1-form, respectively.
Note that boundaries make it necessary to perform a 3+1
split of space-time geometry and physical fields.
As such, they introduce non-covariant elements in the analysis
of problems relevant for quantum gravity. This seems to be 
an unavoidable feature, although the boundary conditions may
be written in a covariant way (see sections 3 and 4).

In the light of (2.2), the boundary conditions
$$
\Bigr[h_{ij}\Bigr]_{\partial M}=0
\eqno (2.3a)
$$
are gauge-invariant, i.e.
$$
\Bigr[{ }^{\varphi}h_{ij}\Bigr]_{\partial M}=0
\eqno (2.3b)
$$
if and only if the whole ghost 1-form obeys homogeneous
Dirichlet conditions, so that
$$
\Bigr[\varphi_{0}\Bigr]_{\partial M}=0
\eqno (2.4)
$$
$$
\Bigr[\varphi_{i}\Bigr]_{\partial M}=0.
\eqno (2.5)
$$
The conditions (2.4) and (2.5) are necessary and sufficient since
$\varphi_{0}$ and $\varphi_{i}$ are independent, and
three-dimensional covariant differentiation commutes with the
operation of restriction at the boundary. Indeed, we are
assuming that the boundary is smooth and not totally geodesic,
i.e. $K_{jl} \not = 0$. However, at those points of
$\partial M$ where the extrinsic-curvature tensor vanishes,
the condition (2.4) is no longer necessary.

The problem now arises to impose boundary conditions on the
remaining set of metric perturbations. The key point is to
make sure that the invariance of such boundary conditions under
the transformations (2.1) is again guaranteed by (2.4) and (2.5),
since otherwise one would obtain incompatible sets of
boundary conditions on the ghost 1-form. Indeed, on using
the Faddeev-Popov formalism for the amplitudes of quantum
gravity, it is necessary to use a gauge-averaging term in
the Euclidean action, of the form 
$$
I_{\rm g.a.} \equiv {1\over 32 \pi G \alpha}
\int_{M}\Phi_{a}(h)\Phi^{a}(h)\sqrt{g} \; d^{4}x
\eqno (2.6)
$$
where $G$ is Newton's constant, 
$\Phi_{a}(h)$ is any gauge-averaging 
functional which leads to self-adjoint elliptic 
(and hence non-degenerate) operators
on metric and ghost perturbations, and $\alpha$ is an arbitrary
dimensionless parameter. As in all our analysis, 
$\sqrt{g}d^{4}x$ is the invariant integration measure with
respect to the background 4-metric. In particular, if the
de Donder gauge is chosen, i.e. (with $a,b=0,1,2,3$)
$$
\Phi_{a}^{dD}(h) \equiv E_{a}^{\; \; b \; cf} \; \nabla_{b}h_{cf}
=\nabla^{b}\Bigr(h_{ab}-{1\over 2}g_{ab}g^{cf}h_{cf}\Bigr)
\eqno (2.7)
$$
where $E^{ab \; cd} \equiv g^{a(c} \; g^{d)b}
-{1\over 2}g^{ab} g^{cd}$, one finds that
$$
\delta \Phi_{a}^{dD} \equiv 
\Phi_{a}^{dD}(h)-\Phi_{a}^{dD}({ }^{\varphi} h)
=-{1\over 2}\Bigr(g_{a}^{\; \; b}\cstok{\ }
+R_{a}^{\; \; b}\Bigr)\varphi_{b}
\eqno (2.8)
$$
where $\cstok{\ } \equiv g^{ab}\nabla_{a}\nabla_{b}$, and
$R_{ab}$ is the Ricci tensor of the background. The operator
$-\Bigr(g_{a}^{\; \; b}\cstok{\ }
+R_{a}^{\; \; b}\Bigr)$ is elliptic and,
of course, acts linearly on the ghost 1-form.
Thus, if one imposes the boundary conditions
$$
\Bigr[\Phi_{0}^{dD}(h)\Bigr]_{\partial M}=0
\eqno (2.9a)
$$
$$
\Bigr[\Phi_{i}^{dD}(h)\Bigr]_{\partial M}=0
\eqno (2.10a)
$$
their invariance under (2.1) is guaranteed when (2.4) and (2.5)
hold, by virtue of (2.8). Hence one also has
$$
\Bigr[\Phi_{0}^{dD}({ }^{\varphi} h)\Bigr]_{\partial M}=0
\eqno (2.9b)
$$
$$
\Bigr[\Phi_{i}^{dD}({ }^{\varphi} h)\Bigr]_{\partial M}=0.
\eqno (2.10b)
$$
Note that the boundary conditions on the ghost 1-form
become redundant if one also imposes the conditions (2.3b),
(2.9b) and (2.10b). Nevertheless, we shall always write them
explicitly, since the ghost 1-form plays a key role
in quantum gravity. 

Of course, the most general scheme does {\it not}
depend on the choice of the de Donder term (see section 5), 
so that it relies on (2.3a), (2.3b), (2.4), (2.5), jointly with 
$$
\Bigr[\Phi_{0}(h)\Bigr]_{\partial M}=0
\eqno (2.11a)
$$
$$
\Bigr[\Phi_{0}({ }^{\varphi} h)\Bigr]_{\partial M}=0
\eqno (2.11b)
$$
$$
\Bigr[\Phi_{i}(h)\Bigr]_{\partial M}=0
\eqno (2.12a)
$$
$$
\Bigr[\Phi_{i}({ }^{\varphi} h)\Bigr]_{\partial M}=0.
\eqno (2.12b)
$$
Again, it is enough to write (2.3a), (2.11a), (2.12a),
(2.4), (2.5), or (2.3a), (2.3b) jointly with (2.11a), (2.11b)
and (2.12a), (2.12b).
\vskip 0.3cm
\leftline {\bf 3. Projection and nilpotent operators}
\vskip 0.3cm
\noindent
Following [8, 9], we study the Barvinsky boundary conditions 
of section 2 for
the semiclassical $\langle {\rm out}|{\rm in} \rangle$ amplitude 
of Euclidean quantum gravity when
a flat four-dimensional background $(M,g)$ is bounded by a
smooth three-dimensional boundary 
$(\partial M,\gamma)$. 
The analysis in arbitrary $d$-dimensional flat manifolds with
smooth $(d-1)$-dimensional boundary can be performed 
along the same lines. 

As the first step in our geometric analysis,
we have to re-express such boundary conditions in
a manifestly covariant way. For this purpose, we 
consider the four-dimensional tensor field $q$ on
$(M,g)$ defined as 
$$
q_{ab} \equiv g_{ab}-n_{a}n_{b}
\eqno (3.1)
$$
whose restriction to $(\partial M,\gamma)$ coincides with the
metric $\gamma_{ij}$ on $\partial M$. Here, $n^{a}$ is the 
inward pointing normal to $\partial M$
with unit norm, i.e. $n_{a}n^{a}=1$. Of course, $q_{ab}$ is
a projector of vector fields onto the surface $\Sigma$ 
orthogonal to the normal vector $n^{a}$, i.e.
$q_{ab}n^{b}=0$.
The boundary conditions (2.3a) are then expressed as
$$
[\Pi \; h]_{\partial M}=0
\eqno (3.2)
$$
where $\Pi$ is a projector of symmetric 2-forms onto 
$\partial M$, defined as
$$
\Pi_{ab}^{\; \; \; cd} \equiv 
q_{\; \; (a}^{c} \; q_{\; \; b)}^{d}.
\eqno (3.3)
$$

In the following we choose the de Donder gauge-averaging functional 
defined in (2.7). 
Given the Levi-Civita connection $\nabla$ of the background,
the introduction of the differential operators $\nabla_{(n)}$
and ${\widetilde \nabla}_{a}$ defined as
$$
\nabla_{(n)} \equiv n^{a}\nabla_{a}
\eqno (3.4)
$$
$$
{\widetilde \nabla}_{a} \equiv q_{\; \; a}^{b} \;
\nabla_{b}
\eqno (3.5)
$$
makes it now possible to write the covariant form of (2.9a)
and (2.10a) as
$$
\Bigr[(A \; \nabla_{(n)}+B^{e} \; {\widetilde \nabla}_{e})
h\Bigr]_{\partial M}=0
\eqno (3.6)
$$
where the matrices $A$ and $B^{e}$ turn out to be 
$$
A_{ab}^{\; \; \; cd} \equiv n_{a}n_{b}\Bigr(n^{c}n^{d}
-q^{cd}\Bigr)
 +2n_{(a} \; q_{\; \; b)}^{(c} \; n^{d)}
\eqno (3.7)
$$
$$ \eqalignno{
B_{ab}^{\; \; \; cd,e} 
&\equiv 2n_{a}n_{b} \; n^{(c} \;q^{d)e}
-n_{(a} \; q_{\; \; b)}^{e}n^{c}n^{d} \cr
&+2n_{(a} \; q^{(c}_{\ b)}q^{d)e}
-n_{(a} \; q^{e}_{\ b)}q^{cd}.
&(3.8)\cr}
$$
Interestingly, a peculiar property of this set of boundary
conditions is that $A$ and $B^{e}$ are not symmetric 
under the interchange of $ab$ and $cd$, and $A$ is {\it not} a
projection operator. By contrast, $\Pi$ is symmetric under the
above interchange, and is a projector by definition. 

One should also bear in mind that, for any $d$-dimensional 
background ($d=4$ in our case), the following property holds:
$$
{\rm rank}(A)+{\rm rank}(\Pi)={d(d+1)\over 2}.
\eqno (3.9)
$$
This condition ensures that the gauge-invariant boundary
conditions (3.2) and (3.6) 
are complete in that they fix all components of
metric perturbations, and do not introduce any spurious
restrictions which would lead to an overdetermined problem.

Note that one can decompose the matrix $A$ in the form
$$
A=\pi+p-\nu
\eqno (3.10)
$$
where the matrices $\pi$, $p$ and $\nu$ are defined by
$$
\pi_{ab}^{\ \ cd} \equiv n_{a}n_{b}\;n^{c}n^{d}
\eqno (3.11)
$$
$$
p_{ab}^{\ \ cd} \equiv 2n_{(a}\;q_{\ b)}^{(c}n^{d)}
\eqno (3.12)
$$
$$
\nu_{ab}^{\ \ cd} \equiv n_{a}n_{b} \; q^{cd}.
\eqno (3.13)
$$
It is easy to see that $\Pi$, $\pi$ and $p$ are projection operators,
i.e.
$$
\Pi^2=\Pi \qquad \pi^2=\pi \qquad p^2=p
\eqno (3.14)
$$
$$
\Pi\,\pi=\pi\,\Pi
=\Pi\, p=p\,\Pi
=\pi\,p=p\,\pi=0
\eqno (3.15)
$$
$$
\Pi+\pi+p=\II
\eqno (3.16)
$$
$\II$ being the identity matrix in the vector space of symmetric
$2$-forms, $\II_{ab}^{\ \ cd}\equiv\delta^c_{\ (a}\delta^d_{\ b)}$,
whereas the matrix $\nu$ is not a projector but a nilpotent  
matrix, i.e.
$$
\nu^{2}=0
\eqno (3.17)
$$
which is orthogonal to $p$
$$
p\,\nu=\nu\,p=0
\eqno (3.18)
$$
whilst
$$
\pi \; \nu= \nu
\eqno (3.19)
$$
$$
\nu \; \pi=0.
\eqno (3.20)
$$
Moreover, the projector $\Pi$ annihilates $\nu$ from the left,
$\Pi\,\nu=0$,
but not in the reverse order, since $\nu\,\Pi=\nu $.
By virtue of (3.17)--(3.20), one has
$$
A \; \nu = \nu
\eqno (3.21)
$$
whilst
$$
\nu \; A = 0.
\eqno (3.22)
$$
In the light of (3.10), (3.16) and (3.17) one sees immediately that
the matrix
$$
\Pi+A=\II-\nu
\eqno (3.23)
$$
is not degenerate and has the inverse
$$
(\Pi+A)^{-1}=\II+\nu .
\eqno (3.24)
$$
Thus,
the action of $A$ and $B^{e}$ on $h$ yields tensor fields which
are orthogonal to $\Pi$, i.e.
$$
\Pi \; A=0
\eqno (3.25)
$$
$$
\Pi \; B^{e}=0.
\eqno (3.26)
$$
On the other hand, $A$ and $B^{e}$ do not commute
with $\Pi$, and hence one finds that
$$
A \; \Pi =-\nu 
\eqno (3.27)
$$
$$
B_{ab}^{\; \; \; cd,e} \; \Pi_{cd}^{\; \; \; fg}
=2n_{(a} \; q_{\; \; b)}^{(f} \; q^{g)e}
-n_{(a} \; q^{e}_{\; b)}q^{fg}.
\eqno (3.28)
$$

By virtue of (3.25) and (3.26), it is possible to express $A$ and
$B^{e}$ as 
$$
A=(\II-\Pi)A
\eqno (3.29)
$$
$$
B^{e}=(\II-\Pi)B^{e}.
\eqno (3.30)
$$
Thus, an equivalent expression of the boundary conditions 
(3.6) is
$$
\Bigr[(\II-\Pi)(A \; \nabla_{(n)}+B^{e} \; {\widetilde \nabla}_{e})
h\Bigr]_{\partial M}=0.
\eqno (3.31)
$$
\vskip 0.3cm
\leftline {\bf 4. Equivalent form of the boundary conditions}
\vskip 0.3cm
\noindent
It is now convenient to transform slightly the 
form of the boundary conditions. This makes it easier to
compare our analysis with previous work in the literature [14],
and can be applied to the geometric analysis of heat-kernel
asymptotics (cf [2--4]). 
For this purpose, let us define the matrix $E=(E_{ab}^{\ \ cd})$ 
with elements
$$
E_{ab}^{\ \ cd} \equiv \delta^c_{\ (a}\delta^d_{\ b)}
-{1\over 2}g_{ab}g^{cd}.
\eqno (4.1)
$$
Substituting here $g_{ab}=q_{ab}
+n_{a}n_{b}$ we obtain the matrix $E$ in the form
$$
E=\II-{1\over 2}(\nu+\nu^T)-{1\over 2}\pi-{1\over 2}V.
\eqno (4.2)
$$
where $T$ denotes the transposition, and the matrix $V$ is defined by
$$
V_{ab}^{\ \ cd} \equiv q_{ab}q^{cd}.
\eqno (4.3)
$$
Now, using the formulae of the previous section, we obtain easily
$$
(\II-\Pi)E=p+{1\over 2}(\pi-\nu).
\eqno(4.4)
$$
It is not difficult to see that this can be 
expressed in terms of the matrix $A$
$$
(\II-\Pi)E={1\over 2}(\II+p)A.
\eqno (4.5)
$$
Noting that
$$
(\II+p)^{-1}=\II-{1\over 2}p
\eqno (4.6)
$$
we find from (4.5)
$$
A=2\left(\II-{1\over 2}p\right)(\II-\Pi)E.
\eqno (4.7)
$$
Therefore, the boundary conditions (3.6) 
(or (3.31)) can be re-written in the form 
$$
\Bigr[(\II-\Pi)\left(E \; \nabla_{(n)}
+{1\over 2}(\II+p)B^{e} \; {\widetilde \nabla}_{e}\right)
h\Bigr]_{\partial M}=0.
\eqno (4.8)
$$
Further we transform the operator 
${\widetilde \nabla}_{e}$ as follows:
$$
{\widetilde\nabla}_{e}
={\widetilde\nabla}_{e}(\II-\Pi)
+{\widetilde\nabla}_{e}\Pi 
=\left((\II-\Pi){\widetilde\nabla}_{e}
-({\widetilde\nabla}_{e}\Pi)\right)(\II-\Pi)
+{\widetilde\nabla}_{e}\Pi.
\eqno (4.9)
$$
Taking into account the boundary condition 
(3.2) on the spatial components of $h$, one finds
$$
[{\widetilde\nabla}_{e} h]_{\partial M}
=\left[\left((\II-\Pi){\widetilde\nabla}_{e}
-({\widetilde\nabla}_{e}\Pi)\right)(\II-\Pi)h\right]_{\partial M}.
\eqno (4.10)
$$
Thus, the boundary conditions take the form
$$
[\Pi \; h]_{\partial M}=0
\eqno (4.11)
$$
$$
\Bigr[(\II-\Pi)\left(E \; \nabla_{(n)}
+F^{e}\; {\widetilde \nabla}_{e}
+\; {\widetilde \nabla}_{e}F^{e}+D\right)
h\Bigr]_{\partial M}=0
\eqno (4.12)
$$
where
$$
F^{e} \equiv {1\over 4}(\II+p)B^e(\II-\Pi)
\eqno (4.13)
$$
$$
D \equiv -{1\over 2}(\II+p)B^{e}({\widetilde\nabla}_{e}\Pi)(\II-\Pi)
-(\II-\Pi)({\widetilde\nabla}_{e}F^{e})(\II-\Pi).
\eqno (4.14)
$$
Using the explicit formulae of the previous 
section for the matrices $B^{e}$ and $\Pi$ one obtains 
$$
F_{ab}^{\ \ \ cd\, ,e}={1\over 2}n_an_bn^{(c}q^{d)e}
-{1\over 2}n_{(a}q_{\ b)}^{e}n^cn^d
\eqno (4.15)
$$
$$
D_{ab}^{\ \ cd}=2n_{(a}q_{\ b)}^{(c}n^{d)}{\rm Tr}K.
\eqno (4.16)
$$ 
It is easy to see that the matrix $D$ is 
proportional to the projector $p$
$$
D=p {\rm Tr}K.
\eqno (4.17)
$$
These boundary conditions are similar to 
the mixed form of generalized boundary
conditions considered in [14]. The geometric theory of heat-kernel
asymptotics resulting from (4.11) and (4.12) remains unknown, and
is a difficult task in Euclidean quantum gravity.

Note that the matrix $F^{e}$ is antisymmetric and the matrix $D$ 
is symmetric with respect to
the interchange of the pairs of indices $ab$ 
and $cd$, i.e.
$$
F^{ab\; cd,e}
=-F^{cd\; ab,e}
\eqno (4.18)
$$
$$
D^{ab\; cd}=D^{cd\;ab}
\eqno (4.19)
$$
and that they are both orthogonal to the projector $\Pi$
$$
F^{e}\Pi=\Pi F^{e}=0
\eqno (4.20)
$$
$$
D\Pi=\Pi D=0.
\eqno (4.21)
$$
\vskip 0.3cm
\leftline {\bf 5. Symmetry of the Laplace operator}
\vskip 0.3cm
\noindent
A crucial point in our analysis is the proof that the boundary
conditions (4.11), (4.12) lead to a self-adjoint operator on
metric perturbations. The de Donder 
gauge-averaging term has the effect of reducing such an operator
to the Laplace operator $-\sq{\ } \equiv -g^{ab}\nabla_{a}
\nabla_{b}$, where $\nabla_{a}$ denotes covariant differentiation 
with respect to the Levi-Civita connection of the background $M$.
As a first step, one has to prove that $\sq{\ }$ is symmetric. 
This means that, denoting by $\eta$ and $h$ any two
elements of the space ${\cal D}(M)$ of $C^{\infty}$, symmetric
tensor fields on $(M,g)$ of type $(0,2)$, and defining (see (4.1))
$$
(\eta,h) \equiv \int_{M}d^{4}x \; \sqrt{g}
<\eta,E \; h>
\eqno (5.1)
$$
where 
$$
<\eta,E \; h> \equiv \eta_{ab}E^{ab\; cd}h_{cd}
\eqno (5.2)
$$
the following property should hold:
$$
I(\eta,h)\equiv (\eta,\sq{\ }h)-(\sq{\ }\eta,h)=0
\eqno (5.3)
$$
for all $\eta,h \in {\cal D}(M)$ and obeying the boundary
conditions (4.11), (4.12).  

In general, the left-hand side of (5.3) takes the form
$$
I(\eta,h)=\int_{\partial M}\biggr[<\eta, E \nabla_{(n)}h>
-<\nabla_{(n)}\eta, E h>
\biggr]
\sqrt{\gamma}d^3 x
\eqno (5.4)
$$
where $\gamma$ is the determinant 
of the 3-metric of the boundary.
In our boundary-value problem,
spatial perturbations $\Pi\eta$ and $\Pi h$ are set to zero
at the boundary (see (4.11)), and hence only the normal components
$(\II-\Pi)\eta$ and $(\II-\Pi) h$ contribute to (5.3). 
Therefore, one has
$$ \eqalignno{
I(\eta,h)&=\int_{\partial M}\biggr[<\eta, 
(\II-\Pi)E \nabla_{(n)}h>\cr
&-<(\II-\Pi)E\nabla_{(n)}\eta, h>
\biggr]
\sqrt{\gamma}d^{3}x.
&(5.5)\cr}
$$
Using now the second boundary condition (4.12) one obtains
$$
I(\eta,h)=\int_{\partial M}\biggr[<\eta, \Lambda h>
-<\Lambda\eta, h>
\biggr]
\sqrt{\gamma}d^{3}x
\eqno (5.6)
$$
where
$$
\Lambda \equiv (\II-\Pi)\left(F^{e}\; {\widetilde \nabla}_{e}
+\; {\widetilde \nabla}_{e}F^{e}+D\right)(\II-\Pi)
\eqno (5.7)
$$
is a first-order differential operator 
on the boundary. Integrating by parts it is 
immediately seen that this operator is symmetric 
$$
\Lambda^{\dag}=\Lambda
\eqno (5.8)
$$
by virtue of the antisymmetry of the matrix 
$F^{e}$ and the symmetry of the matrix $D$. 
Thus, the antisymmetric form $I(\eta,h)$ 
vanishes, and this proves that the Laplacian 
with the boundary conditions (4.11), (4.12) is symmetric.
In a non-covariant analysis, the imposition of the boundary
conditions in the form (1.1)--(1.3) shows that $I(\eta,h)$ 
reduces to the integral over $\partial M$ of the total
divergence $\Bigr[-\eta_{00}h^{0i}
+h_{00}\eta^{0i}\Bigr]_{\mid i}$, and hence vanishes by virtue
of Stokes' theorem (and bearing in mind that 
$\partial \partial M=0$).

The task now remains to prove that self-adjoint extensions exist
and are unique. This appears feasible, since one deals with a
Laplace operator with a Dirichlet sector resulting from (4.11). 
Nevertheless, (5.6) already expresses a non-trivial property:
mixed boundary conditions which are completely invariant under
infinitesimal diffeomorphisms can be consistently imposed
in Euclidean quantum gravity.
\vskip 0.3cm
\leftline {\bf 6. Other choices of mixed boundary conditions}
\vskip 0.3cm
\noindent
The technical problems of section 4 in obtaining the
geometric form of heat-kernel asymptotics result from an involved set
of mixed boundary conditions on the normal components of metric
perturbations. Hence we now study boundary operators whose
action on $h_{00}$ and $h_{0i}$ is instead 
very simple. The first set of boundary conditions is the
covariant version of those analyzed in [16]. They read
$$
\Bigr[n^{b}h_{ab}\Bigr]_{\partial M}=0
\eqno (6.1)
$$
$$
\left[\Bigr(\nabla_{(n)}+{(2+u)\over 3}({\rm Tr}K)\Bigr)
\Bigr(\Pi_{ab}^{\; \; \; cd} \; h_{cd}\Bigr)\right]_{\partial M}=0
\eqno (6.2)
$$
where $u$ is a dimensionless parameter. The non-covariant 
formulation of (6.2) requires that 
${\partial h_{ij}\over \partial \tau}+{u\over \tau}h_{ij}$
should vanish at the boundary. Hence one is dealing with
Robin conditions on $h_{ij}$ [16].
Note that this is {\it not}
the Barvinsky framework. We are still using the de Donder 
gauge-averaging functional, and hence the operator acting on
metric perturbations reduces to $-\cstok{\ }$ in our flat 
Euclidean background. The boundary conditions (6.1) and (6.2) 
represent the extension to gravity of the scheme used in setting
electric boundary conditions for Euclidean Maxwell theory. 
However, unlike Maxwell's theory, they are not completely
gauge-invariant [16]. When $u=0$, (6.2) sets to zero at the boundary
the linearized magnetic curvature, obtained out of the Weyl
tensor [17]. Moreover, the lack of complete gauge invariance
of the boundary conditions implies that, even on the mass
shell, transition amplitudes may depend on the specific form
of the gauge-averaging functional.  

According to the definition (5.1), one thus finds that
the operator $-\cstok{\ }$ is symmetric if and only if the 
following surface integral vanishes:
$$
\eqalignno{
I_{B} &\equiv 
(\eta, \sq h)-(\sq \eta, h) \cr
& =\int_{\partial M} \biggr[\eta^{ij} \; 
\nabla_{(n)}\Bigr(h_{ij}-{1\over 2}g_{ij}{\hat h}\Bigr)
-h^{ij}\nabla_{(n)}\Bigr(\eta_{ij}
-{1\over 2}g_{ij}{\hat \eta}\Bigr)\biggr]
\sqrt{\gamma}d^{3}x
&(6.3)\cr}
$$
where ${\hat h} \equiv g^{ab}h_{ab}, {\hat \eta} \equiv
g^{ab}\eta_{ab}$.
In fact, it is obvious that the boundary conditions (6.1), (6.2)
do satisfy this condition and hence lead to a symmetric Laplace operator,
since the integrand in (6.3) is a linear combination of 
$\eta^{ij}h_{ij}$ and ${\hat \eta} {\hat h}$ 
with vanishing coefficients.

In the Barvinsky framework, the boundary conditions (6.1) may
still be obtained if one uses the axial gauge-averaging
functional $\Phi_{a}^{A}(h) \equiv n^{b}h_{ab}$.
The resulting
ghost operator takes the form
$$
{\cal F}_{a}^{\ b}=(\delta_{a}^{\ b}+n_{a}n^{b})
\nabla_{(n)}+n^{b}{\widetilde \nabla}_{a}
\eqno (6.4)
$$
with Dirichlet boundary conditions (2.4) and (2.5) on the ghost field.
It is not difficult to show that with Dirichlet boundary conditions 
the ghost operator (6.4) does not have any eigenfunctions at all.
Indeed, consider the eigenvalue equation 
$$
{\cal F}\varphi_{\lambda}=\lambda \varphi_{\lambda}.
\eqno (6.5)
$$
The solution of this equation in the coordinates 
$\tau, {\hat x}$ takes the form
$$
{\varphi_{0}}_{\lambda}(\tau,\hat x)
=\exp\left({1\over 2}\lambda\tau\right){f_{0}}_{\lambda}(\hat x)
\eqno (6.6)
$$
$$ \eqalignno{
\varphi_{i\ \lambda}(\tau,\hat x)
&=\exp\left(\lambda\tau\right)
g_{ij}(\tau,\hat x)f^{j}_{\lambda}(\hat x) \cr
&-\int\limits_{0}^{\tau} d y 
\exp\left[\lambda\left(\tau-{1\over 2}y\right)\right]
g_{ij}(\tau,\hat x)g^{jk}(y,\hat x)\hat\nabla_{k} 
{f_{0}}_{\lambda}(\hat x).
&(6.7)\cr}
$$
Now imposing Dirichlet boundary conditions one finds
${f_{0}}_{\lambda}={f^{i}}_{\lambda}=0$, and hence
$\varphi_{\lambda}=0$ for any $\lambda$.
Thus, ghost fields do not contribute 
at all to the transition amplitudes. Note that this is a peculiar
property of Barvinsky boundary conditions. The use of the axial
gauge-averaging functional does not imply, by itself, that the
ghost should vanish identically, unless the whole ghost 1-form is
set to zero at the boundary, as in our case.

As in the previous sections, we impose the boundary conditions 
(3.2) on the spatial components of metric perturbations.
The other components of the field $h_{ab}$
vanish everywhere in the axial gauge
and, of course, at the boundary.
This means, by the way, that {\it all} components of metric 
perturbations vanish at the boundary. Hence
all possible surface terms in the action vanish in this gauge,
and any second-order differential operator 
is in fact symmetric in this particular case.
\vskip 5cm
\leftline {\bf 7. 1-loop divergence in the axial gauge}
\vskip 0.3cm
\noindent
In the absence of boundaries, there is indeed a rich literature
on the axial gauge in quantum gravity and for quantized gauge
theories [18--23]. In [18], the starting point was the analysis
of {\it infrared} properties of quantum gravity in the axial
gauge. It was then shown that gravitons decouple from the
Faddeev-Popov ghosts, and that the leading infrared divergences
exponentiate and vanish in the exponent in the scattering of
gravitons for pure Einstein gravity. This led to a series of
difficult 1-loop calculations, showing that the graviton 
self-energy is non-transverse and $n_{a}$-dependent [19, 20]. 
In [21], all counterterms of quantum gravity were evaluated 
at 1-loop order in the axial gauge, whilst further progress 
for gauge theories was made in [22], and a comprehensive
review appears in [23].

In this section, however, we are interested in the
{\it ultraviolet} divergences of pure gravity in the presence
of boundaries in the axial gauge. The framework under
consideration is relevant for 1-loop quantum cosmology [17] 
and the 1-loop analysis of partition functions in Euclidean
quantum gravity. Thus, unlike [19, 20], we do not study the
graviton self-energy, but we focus on the scaling properties
of 1-loop quantum gravity encoded in the $\zeta(0)$ value [17].
The consideration of the axial gauge is suggested by the 
general scheme of section 2 for diffeomorphism-invariant
boundary conditions, since all metric perturbations are then
set to zero at the boundary in the axial gauge.

We begin our analysis by {\it fixing} the axial gauge by the
Dirac delta in the path integral, i.e. without gauge averaging.
Thus, metric perturbations satisfy the relation
$h_{ab}=\Pi_{ab}^{\ \ cd}h_{cd}$
with $\Pi$ defined in (3.3).
Hence the graviton operator $\Delta_{A}$ in the axial gauge
is obtained by the projection of the operator in the quadratic
part of the action without the gauge-averaging term, i.e.
$S_{2}=\int\limits_{M} d^{4}x \sqrt{g} 
\; {1\over 2} \; h_{ab} \; \Delta^{ab,cd} \; h_{cd}$, as
$\Delta_A=\Pi\Delta\Pi$.
In flat Euclidean space the operator $\Delta$ reduces to the well known 
form [24, 25]
$$ \eqalignno{
\Delta^{ab,cd}&=
-\left(g^{a(c}g^{d)b}-g^{ab}g^{cd}\right)\sq 
-g^{cd}\nabla^{(a}\nabla^{b)}-g^{ab}\nabla^{(c}\nabla^{d)}\cr
&+2\nabla^{(a}g^{b)(c}\nabla^{d)}.
&(7.1)\cr}
$$
One should stress that the graviton operator 
$\Delta_{A}$ in the axial gauge depends,
of course, on the vector $n^{a}$ through the projection operator $\Pi$.
Since in the axial gauge $h=\Pi h$,
the spectrum of the operator $\Delta_{A}$ can be obtained 
by studying the spectrum of the operator $\Delta$ in (7.1)
$$
\Delta_{ab}^{\ \ cd} \; h_{(\lambda)cd}=\lambda h_{(\lambda)ab}
\eqno (7.2)
$$
with the boundary conditions (3.2), or explicitly,
$$
\eqalignno{
\; & -\cstok{\ }h_{(\lambda)ab}
+g_{ab}\cstok{\ }h_{(\lambda)}
-\nabla_{a}\nabla_{b}h_{(\lambda)}
-g_{ab}\nabla_{c}\nabla_{d}h^{cd}_{(\lambda)}
+2 \nabla_{c} \nabla_{(a} \; h_{(\lambda) \; b)}^{c}\cr
&=\lambda h_{(\lambda)\; ab}.
&(7.3)\cr}
$$
If one acts with the covariant differentiation operator
on (7.3) one finds the equation
$$
\lambda \nabla^{a} \; h_{(\lambda)ab}=0
\eqno (7.4)
$$
which implies that, for any $\lambda \not = 0$, metric perturbations
are {\it transverse} in flat 4-space. 
Moreover, the insertion of (7.4) into (7.3), 
jointly with multiplication by $g^{ab}$ and summation over repeated
indices leads to
$$
\biggr(-\cstok{\ }+{1\over 2}\lambda\biggr)h_{(\lambda)}=0.
\eqno (7.5)
$$
It is indeed well known that the spectrum of the Laplace operator
on compact manifolds is bounded from below [26]. Thus, for $\lambda$
greater than a positive constant, the operator 
$-\cstok{\ }+{1\over 2}\lambda$ is positive-definite, and hence (7.5)
implies that metric perturbations are {\it traceless} as well, i.e. $h=0$.

The only technical problems might arise with zero-modes, i.e. 
non-trivial eigenfunctions belonging to vanishing eigenvalues and
satisfying the given boundary conditions. Although we are not (yet)
able to prove a general theorem, we can however point out that,
in the particular (and relevant) case of flat Euclidean 4-space
bounded by a 3-sphere, no non-trivial basis functions exist.
This can be proved by inspection of the mode-by-mode form of the
coupled eigenvalue equations (2.5)--(2.11) of [27], jointly with
equations (2.12) therein, which define the various operators acting
on perturbative modes of the gravitational field. 

Thus, since the ghost field vanishes identically in the axial gauge,
as well as the normal components of $h_{ab}$, whilst $h_{ij}$ is
only transverse-traceless and no non-trivial zero-modes exist, the
resulting $\zeta(0)$ value coincides with the one first obtained
in [28]
$$
\zeta(0)=\zeta_{\rm TT}(0)=-{278\over 45}.
\eqno (7.6)
$$

It is now instructive to outline the calculation when the 
gauge-averaging method is instead used. The axial-gauge
functional modifies the operator (7.1) by the addition of the
term ${1\over \alpha}n^{(a} \; g^{b)(c} \; n^{d)}$. Thus,
covariant differentiation of (7.3), and its contraction with
$g^{ab}$, lead instead to the equations
$$ \eqalignno{
\; & {1\over 2\alpha}\biggr[\Bigr(K_{\; \; b}^{c} \; n^{d}
+K^{cd} \; n_{b} \Bigr)h_{(\lambda)cd}
+n_{b}n^{d}\nabla^{a}h_{(\lambda)ad}\biggr] \cr
&+{1\over 2\alpha}\biggr[({\rm Tr}K)n^{d}h_{(\lambda)bd}
+n^{d}\nabla_{(n)}h_{(\lambda)bd}\biggr]\cr
&=\lambda \nabla^{a}h_{(\lambda)ab}
&(7.7)\cr}
$$
$$
\biggr(-\cstok{\ }+{1\over 2}\lambda \biggr)h_{(\lambda)}
={1\over 2\alpha}n^{c}n^{d}h_{(\lambda)cd}
\eqno (7.8)
$$
subject to the boundary conditions according to which the whole
set of metric perturbations vanishes at the boundary.
Indeed, in the particular case of flat Euclidean 4-space bounded
by a 3-sphere of radius $a$, the unperturbed extrinsic-curvature
tensor $K_{ij}$ is equal to ${1\over a}g_{ij}$, and
$\nabla_{(n)}h_{(\lambda)b0}$ vanishes $\forall b$ on choosing
$n^{a}=(1,0,0,0)$, if $h_{00}=h_{0i}=0$. Thus, a solution of (7.7)
and (7.8) with the boundary conditions described above is compatible
with having $h_{00}=h_{0i}=0$ everywhere, whilst $h_{ij}$ is
transverse-traceless (and hence $h_{ab}$ as well). Moreover, this is
{\it the} solution, since a unique, 
smooth and analytic solution exists of the
quantum boundary-value problem for $h_{ab}$ with homogeneous
Dirichlet conditions at the boundary.
\vskip 0.3cm
\leftline {\bf 8. Concluding remarks}
\vskip 0.3cm
\noindent
Although the choice of boundary conditions is by no means unique
in physics, the request of mathematical consistency may lead to
severe restrictions, and this is indeed the case in Euclidean
quantum gravity. Motivated by 1-loop quantum cosmology [17], 
this paper has studied the mathematical foundations of the boundary
conditions for semiclassical gravity. The four basic properties
one would like to respect are as follows:
\vskip 0.3cm
\noindent
(i) Invariance of the whole set of boundary conditions under
infinitesimal diffeomorphisms on metric perturbations
(see (2.1)).
\vskip 0.3cm
\noindent
(ii) Preservation of the boundary [29].
\vskip 0.3cm
\noindent
(iii) Local nature of the boundary operators. These should
involve zero- and first-order differential operators,
which may or may not represent (complementary) projectors.
\vskip 0.3cm
\noindent
(iv) Symmetry, and possibly essential self-adjointness, of
the differential operators acting on metric perturbations 
and ghost 1-form.

Among the four different schemes studied so far in the
literature [1, 8--10, 16, 30], attention has been focused
in our paper
on Barvinsky boundary conditions [8--10]. These are the only ones 
which require that the gauge-averaging functional $\Phi_{a}$ 
should vanish at the boundary. They provide a framework which is
gauge-invariant by construction, and are local
in that the boundary operators involve first-order or zero-order
differential operators (cf [30]). The first result of our analysis
is that, in the de Donder gauge, which leads to a minimal
operator on metric perturbations, the boundary
operators involve complementary projectors but also a nilpotent
operator. This is a substantial difference with respect to the
scheme proposed in [1], where only projection operators occur in
the boundary conditions. In Euclidean quantum gravity, the
resulting operator on metric perturbations is symmetric.
Such a proof was lacking in the literature (cf [8] and [30]).

We have also shown that
the boundary conditions (2.11a) and (2.12a) are 
compatible with the request (iv) also in 
non-covariant gauges. For example,
on choosing the axial-gauge functional, we have found that
symmetry of the differential operators is immediately obtained
(section 6). Moreover, the resulting 1-loop divergence has been
found to coincide with the one resulting from transverse-traceless
modes only [28]. This is a non-trivial property, since a gauge
has been found such that the contributions of ghost and gauge
modes vanish separately in the presence of boundaries.
This property is not shared by other non-covariant gauges, 
e.g. the Coulomb gauge for Euclidean Maxwell theory [31], 
where the ghost and gauge contributions cannot be made to vanish
separately for problems with boundary.
Note however that non-covariant choices, like the axial 
gauge $n^{b}h_{ab}=0$, might restrict the class
of background four-geometries to those for which the singularity 
at the origin is avoided (e.g. the 
so-called two-boundary problem [7]), so
that normal components of metric perturbations are well defined.

Last, we have put on solid ground the proof of
symmetry of the graviton operator when the boundary conditions
studied in [16] are imposed. It now remains to be seen whether
such an operator is essentially self-adjoint (cf [32]), and
whether the semiclassical theory is consistent despite the lack
of complete gauge invariance of the boundary conditions
(cf [1, 29, 30]). The former task appears easier, since one
deals with a Laplace-like operator (in flat space) with 
Dirichlet and Robin sectors.

At a technical level, the outstanding open problem is now to
find a geometric theory of heat-kernel asymptotics 
corresponding to the form (4.11) and (4.12) of Barvinsky
boundary conditions in the de Donder gauge. The scheme is 
(far) more involved than the one considered in [2--4], 
since both normal and tangential derivatives of
metric perturbations occur in the boundary conditions.
However, such a step should be undertaken to complete the
geometric description of ultraviolet divergences at 1-loop
on manifolds with boundary.

Last, but not least, one has to prove essential self-adjointness
[32] of the operator acting on metric perturbations when
Barvinsky boundary conditions [10] in linear covariant gauges
are imposed in the case of flat or curved four-dimensional
backgrounds.
\vskip 0.3cm
\leftline {\bf Acknowledgments}
\vskip 0.3cm
\noindent
Anonymous referees made comments which led to a substantial
improvement of the original manuscript, and
we are indebted to Hugh Osborn for sending us a copy of [14].
The work of I Avramidi was supported by the Alexander von Humboldt
Foundation, by the Istituto Nazionale di Fisica Nucleare 
and by the Deutsche Forschungsgemeinschaft.
He is also grateful to the
Naples Section of the INFN for hospitality. The work of A
Kamenshchik was partially supported by the Russian Foundation
for Fundamental Researches through grant No 96-02-16220-a, and
by the Russian Research Project ``Cosmomicrophysics".
\vskip 0.3cm
\leftline {\bf References}
\vskip 0.3cm
\item {[1]}
Luckock H C 1991 {\it J. Math. Phys.} {\bf 32} 1755
\item {[2]}
Moss I G and Poletti S J 1994 {\it Phys. Lett.} 
{\bf 333B} 326
\item {[3]}
Vassilevich D V 1995 {\it J. Math. Phys.} {\bf 36} 3174
\item {[4]}
Gilkey P B 1995 {\it Invariance Theory, the Heat Equation and
the Atiyah-Singer Index Theorem} (Boca Raton: CRC Press)
\item {[5]}
D'Eath P D and Esposito G 1991 {\it Phys. Rev.} D {\bf 43} 3234
\item {[6]}
Kamenshchik A Yu and Mishakov I V 1993 {\it Phys. Rev.}
D {\bf 47} 1380 
\item {[7]}
Esposito G, Kamenshchik A Yu, Mishakov I V and Pollifrone G 1994
{\it Class. Quantum Grav.} {\bf 11} 2939
\item {[8]}
Esposito G, Kamenshchik A Yu, Mishakov I V and Pollifrone G 1995
{\it Phys. Rev.} D {\bf 52} 3457
\item {[9]}
Esposito G 1996 in {\it Quantum Field Theory Under the Influence
of External Conditions} ed M Bordag (Stuttgart-Leipzig: Teubner)
\item {[10]}
Barvinsky A O 1987 {\it Phys. Lett.} {\bf 195B} 344
\item {[11]}
Strocchi F 1993 {\it Selected Topics on 
the General Properties of Quantum Field Theory}
(Singapore: World Scientific)
\item {[12]}
Esposito G, Gionti G, Kamenshchik A Yu, Mishakov I V and
Pollifrone G 1995 {\it Int. J. Mod. Phys.} D {\bf 4} 735
\item {[13]}
Barvinsky A O, Kamenshchik A Yu and Karmazin I P 1992
{\it Ann. Phys.,} {\it NY} {\bf 219} 201
\item {[14]}
McAvity D M and Osborn H 1991 {\it Class. Quantum Grav.}
{\bf 8} 1445
\item {[15]}
York J W 1986 {\it Found. Phys.} {\bf 16} 249
\item {[16]}
Esposito G and Kamenshchik A Yu 1995 {\it Class. Quantum Grav.}
{\bf 12} 2715
\item {[17]}
Esposito G 1994 {\it Quantum Gravity, Quantum Cosmology and
Lorentzian Geometries} ({\it Lecture Notes in Physics}
{\bf m12}) (Berlin: Springer)
\item {[18]}
Matsuki T 1979 {\it Phys. Rev.} D {\bf 19} 2879
\item {[19]}
Capper D M and Leibbrandt G 1982 {\it Phys. Rev.} D {\bf 25}
1009
\item {[20]}
Capper D M and Leibbrandt G 1982 {\it Phys. Rev.} D {\bf 25}
2211
\item {[21]}
Matsuki T 1985 {\it Phys. Rev.} D {\bf 32} 3164
\item {[22]}
Kreuzer M, Piquet O, Rebhan A and Schweda M 1986 
{\it Phys. Lett.} {\bf 169B} 221
\item {[23]}
Leibbrandt G 1987 {\it Rev. Mod. Phys.} {\bf 59}
1067
\item {[24]}
Fierz M and Pauli W 1939 {\it Proc. R. Soc. London} {\bf A173} 211
\item {[25]}
Avramidi I G 1991 {\it Int. J. Mod. Phys.} A {\bf 6} 1693
\item {[26]}
Chavel I 1984 {\it Eigenvalues in Riemannian Geometry}
(New York: Academic Press)
\item {[27]}
Esposito G, Kamenshchik A Yu, Mishakov I V and Pollifrone G 
1994 {\it Phys. Rev.} D {\bf 50} 6329
\item {[28]}
Schleich K 1985 {\it Phys. Rev.} D {\bf 32} 1889
\item {[29]}
Luckock H C and Moss I G 1989 {\it Class. Quantum Grav.}
{\bf 6} 1993
\item {[30]}
Marachevsky V N and Vassilevich D V 1996 
{\it Class. Quantum Grav.} {\bf 13} 645
\item {[31]}
Esposito G and Kamenshchik A Yu 1994 {\it Phys. Lett.} 
{\bf 336B} 324
\item {[32]}
Esposito G, Morales-T\'ecotl H A and Pimentel L O 
1996 {\it Class. Quantum Grav.} {\bf 13} 957

\bye